\documentclass[preprint]{aastex}
\usepackage{epsfig}
\slugcomment{To appear in The Astrophysical Journal}

\begin{document}
\title{Peculiar Extended X-ray Emission around the ``Radio-Loud'' 
Black Hole Candidate 1E1740.7-2942}

\author{Wei Cui\altaffilmark{1}\altaffilmark{2}, N.S. Schulz\altaffilmark{1}, 
F.K. Baganoff\altaffilmark{1}, M.W. Bautz\altaffilmark{1},
J.P. Doty\altaffilmark{1},
G.P. Garmire\altaffilmark{3}, I.F. Mirabel\altaffilmark{4}\altaffilmark{5},
G.R. Ricker\altaffilmark{1}, L.F. Rodr\'\i guez\altaffilmark{6}, 
and S.C. Taylor\altaffilmark{1}} 

\altaffiltext{1}{Center for Space Research, Massachusetts Institute 
of Technology, Cambridge, MA 02139}

\altaffiltext{2}{Department of Physics, Purdue University, 
West Lafayette, IN 47907}

\altaffiltext{3}{Department of Astronomy and Astrophysics, 
Pennsylvania State University, University Park, PA 16802}

\altaffiltext{4}{Service d'Astrophysique, Centre d'Etudes de Saclay,
91191 Gif-sur-Yvette, France}

\altaffiltext{5}{Instituto de Astronom\'\i a y F\'\i sica del Espacio,
Argentina}

\altaffiltext{6}{Instituto de Astronomia, UNAM, Apartado Postal
70-264, 04510 M\'{e}xico, DF, Mexico}

\begin{abstract}

We present the discovery of peculiar extended X-ray emission around
1E1740.7-2942, a black hole candidate that is known to produce
prominent, persistent radio jets. The data was obtained with the 
High-Energy Transmission Grating Spectrometer (HETGS) aboard the 
{\em Chandra X-ray Observatory}. The zeroth-order image reveals 
an elongated feature about 3\arcsec\ in length that is roughly 
{\em perpendicular} to the radio lobes (or jets). The feature 
is roughly symmetric about the point source. It is spatially 
resolved in the long direction but not in the short direction. 

The position of 1E1740.7-2942 was determined with a statistical 
accuracy of $\sim$0.06\arcsec\ in the right ascension and 
$\sim$0.04\arcsec\ in the declination, thanks to {\em Chandra's} 
unprecedented spatial resolution. It is about 0.6\arcsec\ from the 
radio position but the difference is well within the uncertainty in 
the absolute aspect solutions of the observation. The dispersed HETGS 
spectra of 1E1740.7-2942 show evidence for the presence of weak, 
narrow emission lines, although the statistics are quite limited. 
We discuss possible origins of the extended emission and the 
implications of the emission lines.

\end{abstract}

\keywords{binaries: general --- stars: individual (1E1740.7-2942) ---
X-ray: stars}
 
\section{Introduction}

Discovered by the {\em Einstein} satellite (Hertz \& Grindlay 1984),
1E1740.7-2942 appears as a persistent X-ray source in the sky, only 
about 50\arcmin\ away from the center of the Galaxy. It is in fact
the brightest persistent source at high energies ($\gtrsim$ 30 keV) 
within a few degrees of the Galactic center (e.g., Skinner et al. 
1987; Sunyaev et al. 1991). At low energies ($\lesssim$ 3 keV), 
however, the X-ray emission from the source is strongly attenuated 
by the intervening matter along the line of sight. The inferred 
hydrogen column density is around $10^{23}\mbox{ }cm^{-2}$ (see Sheth 
et al. 1996, Churazov et al. 1996, and Sakano et al. 1999 for most 
recent measurements), which is unusually high even for sources 
physically located near the Galactic center. Additional absorption 
can be attributed to the material associated with a molecular cloud 
in which the source is likely embedded (Bally \& Leventhal 1991; 
Mirabel et al. 1991; Phillips et al. 1995; Yan \& Dalgarno 1997).

1E1740.7-2942 has a very hard X-ray spectrum (measured up to $\sim$300 
keV; e.g., Skinner et al. 1991; Cook et al. 1991; Kuznetsov et al. 
1997), which is characteristic of black hole candidates (BHCs) such as 
Cyg X-1. The source also shows strong aperiodic and quasi-periodic 
X-ray variability (Smith et al. 1997; Lin et al. 2000), again similar 
to BHCs. The observed X-ray properties, therefore, make 1E1740.7-2942 
a BHC, although dynamical evidence for such a candidacy is still
lacking. The high foreground extinction precludes the identification 
of the optical counterpart. Extensive searches for the infrared 
counterpart have also been unsuccessful (e.g., Leahy et al. 1992;
Marti et al. 2000). However, 1E1740.7-2942 shows up prominently at
radio wavelengths (Mirabel et al. 1992). In fact, the most remarkable 
property known of the source is the presence of persistent, two-sided 
radio jets, reminiscent of extragalactic radio-loud quasars. Such 
comparisons support the notion that 1E1740.7-2942 is an X-ray binary 
that contains an accreting black hole.

Shortly after its discovery, 1E1740.7-2942 generated a tremendous
amount of excitement for being a possible source of electron-positron
annihilation radiation (Bouchet et al. 1991; Sunyaev et al. 1991),
which had been detected from the general direction of the Galactic 
center by earlier balloon-borne and space-borne experiments (see 
Harris 1997 for a historical account). Suggestions were subsequently 
made that the jets are perhaps made of electron-positron pairs ejected 
from the central black hole at relativistic speeds (Mirabel et
al. 1992) and that the particles will eventually be slowed down and 
annihilated in the surrounding cold, dense molecular cloud. However, 
the reality of 1E1740.7-2942 being the ``great annihilator'' in the 
Galaxy has been brought into fierce debate, mostly because of the 
failure of detecting any annihilation features by the instruments 
aboard the {\em Compton Gamma-ray Observatory}, even in some of the 
simultaneous or contemporaneous observations with those of 
{\em SIGMA/GRANAT} on which the claims of detections were based
(see review by Harris 1997). This remains a highly controversial 
issue.

To shed more light on the unusual environment surrounding
1E1740.7-2942, we observed the source with the {\em Chandra X-ray 
Observatory} (Chandra; Weisskopf \& O'Dell 1997), taking advantage of 
the unprecedented spatial resolution that {\em Chandra} offers (van
Spybroeck et al. 1997). In this paper, we report the discovery of 
peculiar extended X-ray emission around 1E1740.7-2942, and present 
the first high-resolution X-ray spectra of the source.

\section{Observation}

1E1740.7-2942 was observed for about 10 ks on September 25, 1999 with 
the High-Energy Transmission Grating Spectrometer (HETGS; Canizares et 
al. 2000) on board {\em Chandra}, shortly after the Orbital Activation 
and Checkout phase of the mission. The grating spectra are read out by 
the spectroscopic array of the Advanced CCD Imaging Spectrometer
(ACIS; Garmire et al. 2000) at the focal plane of the X-ray
telescope. For the observation, we adopted an alternating exposure 
data mode with three primary (or long) frames (with a readout time of 
3.3 s) followed by one secondary (or short) frame (with a readout time
of 0.3 s). The reason for using the short frames is to obtain some 
zeroth-order data that is essentially free of the effects of
``pile-up'' (i.e., multiple photons hit the same event detection cell 
of the ACIS within one readout frame). This is important for
observations of bright 
sources, such as 1E1740.7-2942, where serious pile-up is expected to 
distort the zeroth-order image. Although an unpiled-up spectrum can 
be obtained from the higher-order grating data in this case, such data 
is of little use for imaging analysis. However, the use of alternating 
exposure modes reduces the observing efficiency. In our case, the 
effective exposure time is only about 6.7 ks and 0.2 ks for the 
long-frame data and short-frame data, respectively.

\section{Data Analysis and Results}

We received data products of the standard processing pipeline from 
the {\em Chandra} X-ray Center (CXC), in which the long-frame data 
and short-frame data were conveniently separated. We checked the 
(Level 2) products against all known caveats (listed at the CXC web 
site) for possible problems and found none (except for a known offset 
in the aspect solutions; see below). We then analyzed the Level 2 
data with the {\em Chandra} Interactive Analysis of Observations 
software (version 1.1) which we obtained from the CXC. For some 
aspects of the spectral analysis we also used custom software as well 
as XSPEC version 10.0.

As expected, the global image (not shown) shows a compact X-ray 
emission region in the zeroth-order data and a characteristic ``X'' 
pattern of higher-order (or dispersed) spectra from two grating 
assemblies, the High Energy Grating (HEG) and Medium Energy Grating 
(MEG). The intrinsic energy resolution of the ACIS is relied upon for  
sorting out the overlapping orders of the dispersed spectra. In 
addition, a faint trace is visible in the 
long-frame image that runs through the zeroth-order image along the 
direction of frame transfer. This is an artifact resulting from 
photons hitting the CCD chip during the frame transfer operation. 
Therefore, such a ``readout trace'' is present only in the 
observations of bright X-ray sources. No other point sources were 
significantly detected over the entire field of view (roughly 
8\arcmin\ $\times$ 50\arcmin).  

\subsection{Extended X-ray Emission}

Zooming in on the zeroth-order image (shown in Fig. 1), we see that
the innermost emission region appears to be elongated in the direction 
of northeast and southwest (but mostly east and west). In Fig. 1,
we displayed an image made from the short-frame data and overlaid it 
with contours from the long-frame data. The image looks reasonably 
smooth and clearly shows a maximum at, presumably, the location of 
the point source. In contrast, the contours appear quite ``lumpy'' and 
actually show a local {\em minimum} at the source position. This is 
almost certainly due to the 
distortion of point spread function (PSF) caused by significant 
pile-up. The pile-up effects are obviously most severe at the position
of the source and can cause an apparent ``valley'' there in the X-ray 
image. The effects can be crudely quantified by deriving the 
{\em expected} count rate of source in the zeroth-order image from
that in the higher-order image, since the latter does not suffer from 
any pile-up. We determined that the expected zeroth-order count rate
of the source should be about 1.9 count/s (or 6.3 count/frame), which 
explains the severe distortion of 
the PSF observed in the long-frame image. Actually, the situation is 
made worse by the fact that 1E1740.7-2942 has an unusually hard spectrum, 
which makes it more likely to cause additional ``lost and undetected''
events (Allen et al. 1998). Indeed, the measured count rate (from the 
zeroth-order image) is only about 0.03 count/s (compared to about 
0.2 count/s for the extended emission surrounding the source). For 
the short-frame data, we estimated 
the count rate by adding up all the charge in the 3x3 event detection 
island around the central pixel in the 0th order image and determined 
a count rate of 1.28 count/s (or 0.38 count/frame), which corresponds 
to a $\sim$30\% pile-up (compared to the expected 1.9 count/s). For 
comparison, the measured count rate of the surrounding extended 
emission is about 0.3 count/s (which is significantly larger than
that from the long-frame data, again indicating severe pile-up 
effects in that case). 
To quantify the effects of lost-and-undetected events for an unusually
hard source like 1E1740.7-2942, we also ran PIMMs to estimate the 
pile-up rate. We obtained a significantly lower value, $\sim$20\%, 
which can clearly be attributed to the fact that PIMMS does not take 
into account the lost-and-undetected events. Even at a pile-up rate
of 30\%, little distortion of the PSF is expected, based on the
results from pre-flight ground calibration tests (Allen et 
al. 1998). However, the short-frame image suffers from the lack of 
statistics, due to the short exposure, which is why it does not 
show the faint ``halo'' emission (only about 0.04 count/s) 
surrounding the point source
and the elongated feature (see outer contours in Fig.~1).

To quantify the elongation seen in the zeroth-order image, we ran
MARX simulations (Wise et al. 1999) for the HETGS 0th order image of 
a point source which included the observed level of background. In
Fig.~2., we fit the simulated MARX PSFs to the short-frame count 
profiles collapsed along the elongation and across it. The measured 
profile (solid line) across the elongation (top of Fig. 2) cannot 
be distiguished from a point source at a confidence level of 
99.9$\%$. Here we determined the full-width-at-half-maximum (FWHM) 
of the profile to be 0.83\arcsec\ (please note, the pointing of the 
observation was slightly off-axis to avoid the CCD node boundary). 
At roughly the same confidence level, we can exclude the possibility 
that the measured profile along the elongation is point-like. The 
width of the elongation is 1.35\arcsec\ FWHM. 

Looking at Fig.~1 more carefully, the long-frame contours seem to
indicate the presence of another region of extended emission, which 
lies to the south of the main feature. Given the severe distortion
of the image by pile-up effects, however, it is entirely possible 
that this and the main feature are the integral part 
of the same extended emission. The smoothness of the short-frame 
image appears to support that. In addition, the fainter and more
extended emission at the outer part of the image is statistically
significant. It is approximately spherically symmetric about the 
point source and can be measured up to about 4\arcsec\ away from 
the source. 

\subsection{Position of 1E1740.7-2942}

We derived the position of the point source in two ways. First, we
constructed the zeroth-order count profiles along the right ascension 
(RA) and declination (Dec) from the the un-distorted short-frame image 
(in sky coordinates). We fitted the profiles with a simple Gaussian 
function to determine the position of the peaks. Although Gaussian 
functions are not a very good representation of the overall count 
profiles (especially around the wings), they are used only to fit a
limited range around the peak of the profiles for determing the peak
position. This allows us to measure the position to a sub-pixel
accuracy, which is only limited by the signal-to-noise ratio (S/N) 
of the short-frame data.

Second, we utilized the fact that the point source should be at the
intersection point of the readout trace and the dispersed MEG and HEG
images. In this case, the problem of deriving the source position 
is over constrained, which allows a more realistic estimate of the
uncertainties (including all systematic effects). We started by fitting
a Gaussian function to the count profile across the readout trace (in
the long-frame image) to determine its center. This gives the position 
of the source along one direction (nearly along the Dec) in sky 
coordinates. We then determined the 
position of the intersection point between the readout trace and 
the MEG or HEG images. The average of the two positions was used as 
the position of the source along the orthogonal direction (nearly
along the RA). The difference between the them provides an estimate of 
the uncertainty in the source position derived using this technique. 

For comparison, the results from both methods are summarized in Table~1. 
It is worth noting that the first method produced slightly smaller error 
bars, in spite of poorer statistics of the short-frame data. This can 
be understood in terms of systematic uncertainties in the dispersed 
HETGS traces and the faintness of the readout trace. 
We also note that there 
is a known offset ($\approx$2.09\arcsec) in the aspect solutions for 
our observation (CXC web page). The numbers shown in the table have
already been corrected for the offset. It is reassuring that the two
methods are in an excellent agreement. 

\subsection{X-ray Spectra}

We constructed the first-order ($\pm 1$) spectra of 1E1740.7-2942 from 
the MEG and HEG data. The higher-order data is ignored because of poor
statistics. The individual spectra are compared for identifying common
features which may only be marginal statistically in each case. The 
general agreement between the two MEG or HEG spectra or between
the MEG and HEG spectra is good, although the statistics of the HEG 
data is poorer. To improve S/N, we first added the plus and minus
orders for the MEG and HEG data separately; we then co-added the MEG
and HEG first orders, with special care in deriving the wavelength
scale for each spectrum. Fig.~3 shows the result. The 
spectrum was binned to $\sim$0.02 \AA\  (which is about the energy 
resolution of the MEG). Note that there are hardly any counts above 
$\sim$5.5 \AA\ , because the source is known to be heavily absorbed. 
A crude fit to the continuum with a simple power law plus absorption 
yielded a hydrogen column density 
$11.8 \pm 0.6 \times 10^{22}\mbox{ }cm^{-2}$ and a photon index $0.90 \pm
0.15$, in reasonable agreement with the ASCA results (Sheth et
al. 1996; Churazov et al. 1996; Sakano et al. 1999). The integrated 
observed 2-10 keV flux is 
$2.2\times 10^{-10}\mbox{ }ergs\mbox{ }s^{-1}\mbox{ }cm^{-2}$. The
flux error is of the order of 10$\%$, which is the current status
of calibration in the HETGS at high energies. The best-fit continuum
model is also shown in Fig. 3 (in solid histograms) but is arbitrarily
shifted down by 15\% to accentuate possible line features (see below).

Deviations from the continuum model are quite apparent, mostly in the 
form of weak, narrow emission lines (though detected only at a 
significance level of roughly $2 \sigma$; see Table~2). The feature 
at $\sim$1.74 \AA\  
(or $\sim$7.1 keV) appears to be an absorption edge, perhaps due to 
cold Fe atoms associated with the interstellar medium. The feature 
centered at $\sim$1.78 \AA\  
(or $\sim$6.96 keV) is probably due to the recombination line of 
H-like Fe ions (Fe XXVI). Note that the feature appears quite broad.
Curiously, there appears to be two lines at 1.70 \AA\  and 1.96 \AA, 
respectively, which cannot be easily identified with any candidate 
elements. We speculate that these lines might also originate in the 
Fe XXVI recombination line but be blue- and red-shifted by the 
approaching and receding jets, respectively. It is worth pointing out 
that in this scenario the asymmetry of the lines about the rest-frame 
energy (see Fig.~3) would be expected because of the transverse
Doppler shift. If the interpretation is correct, the observed radial 
and transverse Doppler shifts would imply that the velocity of the 
jets is about 0.26 c and the inclination angle (with respect to the 
line of sight) about 60\arcdeg, assuming the approaching and receding 
jets are intrinsically identical. Applying the scenario to the
remaining line candidates, we found that about 90\% of them could be
identified with H-like or He-like ions of Ca, S, and Ar and that the
inferred blue- and red-shifted counterparts would give roughly the 
same jet parameters. Table~2 shows a list of the emission lines that
are marginally detected, along with possible identifications, as 
well as the proposed Doppler-shifted counterparts of the lines. In 
contrast, we could only manage to identify 
about 40\% of the lines without invoking any Doppler effects. Note 
that for the line candidates we only consider those that are present 
in at least three out of four first-order ($\pm 1$) MEG and HEG 
spectra. Any real spectral features in the HEG spectra should in 
principle always be present in the MEG spectra because of the 
improved statistics. The only exception is that a feature could fall 
right in the gap between two CCD chips along the MEG arm of the ``X'' 
pattern.

The zeroth-order spectrum could, in principle, shed some light on the
spectral properties of the extended emission. Unfortunately, severe
pile-up causes significant distortion of the spectrum obtained from
the long-frame data, while the short-frame data is not very useful 
due to its poor S/N. We could not tell with sufficient confidence 
whether there are any emission lines that are not present in the 
dispersed spectra (and thus would be associated with the extended 
emission). For comparison, when we binned the observed grating
spectrum (Fig.~3) to the ACIS energy resolution (e.g., 0.05 \AA at 
1.85 \AA), all features vanish, except for a merge between the 
ones at 1.71 \AA and 1.78 \AA. We chose not to pursue the 
subject any further in this work. 

\section{Discussion and Conclusions}

For the first time, {\em Chandra} made it possible to image a possible 
black hole jet system, 1E1740.7-2942, with sub-arcsecond resolution in 
X-rays. Our primary motivation for conducting the observation was to 
search for X-ray emission from the jets themselves, both by direct 
imaging and by searching for Doppler-shifted emission lines as in the 
well-known case of SS~433 (Kotani et al. 1996). 

We found no evidence for any jet emission in the X-ray image. Instead, 
we discovered a region (or regions) of extended emission. For
comparison, a radio image of the system is shown in Fig.~4, with
the inset showing an X-ray image of the region around the radio
core that is overlaid by radio contours with comparable resolution.
We can see that the main X-ray feature is elongated in the direction 
roughly orthogonal to the axis of the radio lobes. The extent of the 
elongation is about 3\arcsec, which corresponds to a 
linear size of $\sim$0.12 pc, assuming that the distance to the 
source is 8.5 kpc. It seems, therefore, much too large to be associated
with an accretion disk. We note that similar features have been
detected in a recent VLBA+VLA observation of SS~433 (Paragi et
al. 2000), although the extended radio emission is about a factor of
30 smaller than the main X-ray feature seen here in 1E1740.7-2942. 
It was speculated that the extended radio features in SS~433 might 
originate in the non-thermal radiation from relativistic electrons 
accelerated by shocks that are formed due to the interaction between 
the ejected matter from the accretion disk and the ambient medium 
(Paragi et al. 2000). Such an ``excretion disk'' scenario (through the
L2 Lagrangian point) was thought to be possible for SS~433 during
episodes of enhanced mass accretion. Perhaps, this scenario is also 
relevant for the much larger extended feature in 1E1740.7-2942. In
this case, the extended X-ray emission observed could be due to the 
reprocessed radiation from the outflowing material that is illuminated 
by the central X-ray source. Alternatively, there might exist an
equatorial wind from the companion star which produces the extended 
X-ray emission in a similar manner. In either case, fluorescent lines
would be expected in the spectrum of the extended emission. 
Unfortunately, our data is not of sufficient quality to verify
that. 

On the other hand, it is perhaps worth asking whether the extended 
X-ray emission could be associated with a previously unknown young 
supernova remnant (SNR). To see whether such a small SNR (of radius 
only ~0.06 pc) is plausible, we adopt the Sedov-Taylor similarity 
solution (e.g., Osterbrock 1989):
\begin{equation}
R_s = 2 t_{100}^{2/5} E_{51}^{1/5} N_0^{-1/5}\mbox{ }pc,
\end{equation}
where $R_s$ is the radius of the shock front, $t_{100}$ is the time
since the explosion in units of 100 years, $E_{51}$ is the initial
kinetic energy in units of $10^{51}$ erg, and $N_0$ is the number 
density of the ambient medium. Clearly, for any reasonable $t_{100}$ 
and $E_{51}$, an enormous ambient density would be required to produce 
a SNR of the size of the observed extended X-ray feature. Besides,
1E1740.7+2942 and its vicinity has been observed with VLA, and no 
SNR has ever been detected (or reported). Therefore, we believe that 
the SNR scenario can be safely ruled out.

Although the identification of the radio counterpart of 1E1740.7-2942
is reasonably firm (Mirabel et al. 1993), there were always lingering
doubts about the association of the X-ray source with the compact
radio core. The sub-arcsecond spatial resolution of {\em Chandra} has 
now allowed us to measure the position of the X-ray source to an
accuracy $\sim$0.06\arcsec. The results (see Table 1) show that the 
X-ray source is about 0.6\arcsec\ east and 0.1\arcsec\ north of the 
compact radio core position. The difference between the X-ray and 
radio measurements is well within the uncertainty in 
the absolute aspect solutions ($<$ 1\arcsec; CXC web page). Given
that in this case only one X-ray source is significantly detected over 
an approximately 8\arcmin\ $\times$ 50\arcmin\ field of view, the 
probability of the source to fall, by chance, to within 
$\sim$0.6\arcsec\ of the radio core is $\sim 8\times 10^{-7}$. It is
worth emphasizing that such a bright source would have been 
detectable {\em anywhere} inside the field of view. Therefore, we 
conclude that it is highly unlikely that 1E1740.7-2942 was 
mis-identified at the radio wavelength.

The high-resolution HETGS spectrum of 1E1740.7-2942 shows evidence for
emission lines (but only at roughly $2 \sigma$ levels). The lines
appear to be intrinsically narrow and weak, so they could have escaped 
the detections by previous X-ray
spectrometers of inferior energy resolution (e.g., the GIS and SIS 
aboard ASCA; Sheth et al. 1996 and Churazov et al. 1996). The lines 
could be identified with H-like or He-like ions of Fe, Ca, S, and Ar. 
This would imply that the line-emitting material is highly ionized, 
which is perhaps not so surprising for an X-ray binary. The lines 
might originate from the accretion disk or the base of the jets where 
material moves relatively slowly. There is also evidence for the 
emission from the jets themselves, which is manifested in the pairs of 
Doppler red- and blue-shifted lines, although it is a bit speculative 
at present given the quality of the data. It is interesting to note, 
however, that every pair of lines would indicate approximately the 
same amount of Doppler shifts and that the inferred velocity and
inclination angle of the twin jets are not unreasonable for this 
system (see \S~3.3; compared to SS~433). A definitive confirmation or 
rejection of the proposed scenario awaits data of much improved S/N 
from future {\em Chandra} observations.

The faint, spherically symmetric halo emission is likely due to the 
scattering of X-rays from 1E1740.7-2942 by interstellar dust. Such
scattering halos are known to exist around many X-ray sources and 
have proven useful for studying the spatial and size distributions of 
the dust grains in our Galaxy (e.g., Mauche \& Gorenstein 1986;
Predehl \& Schmitt 1995). {\em Chandra} now makes it possible, for the 
first time, to directly image the phenomenon. The observations should, 
in principle, allow us to derive spatially resolved energy
distribution of photons in the halo and thus to test theoretical 
models at a more detailed level. Unfortunately, our observation does 
not provide sufficient statistics to warrant such an investigation.

\acknowledgments

We are grateful to Paul Plucinsky and his colleagues at CXC for
clarifying several critical issues regarding the integrity of the 
pipeline products for our observation. We also thank Nancy Evans and
her colleagues at the CXC help desk for providing a definitive answer 
to our inquiry regarding the uncertainty in the absolute aspect 
solutions. This work was supported in part by NASA through contract 
NAS8-38252. NSS and SCT wish to acknowledge
support from NASA through the Smithonian Astrophysical Observatiory 
contract SAO SV1-61010, and IFM support from Conicet/Argentina.

\clearpage
\begin{deluxetable}{lcc}
\tablecolumns{3}
\tablewidth{0pc}
\tablecaption{Position of 1E1740.7-2942\tablenotemark{1}}
\tablehead{
\colhead{Method} & \colhead{RA (J2000)} & \colhead{Dec (J2000)}
}
\startdata
1: PSF Fit  &  $17^h\mbox{ }43^m\mbox{ }54^s.876 \pm 0.004$ &
$-29$\arcdeg\ $44$\arcmin\ $42$\arcsec$.48 \pm 0.04$ \\
2: HETG Fit &  $17^h\mbox{ }43^m\mbox{ }54^s.878 \pm 0.005$ &
$-29$\arcdeg\ $44$\arcmin\ $42$\arcsec$.53 \pm 0.09$ \\
\hline \hline
Radio Position\tablenotemark{2} & $17^h\mbox{ }43^m\mbox{ }54^s.83$ &  
$-29$\arcdeg\ $44$\arcmin\ $42$\arcsec.60
\tablenotetext{1}{The uncertainties shown represent roughly 90\% 
confidence limits, except for those on the RA from the second 
method (see text). }
\tablenotetext{2}{The radio coordinates were derived from an 
expanded dataset, as shown in Fig.~4, and the position is slightly 
different from those given by Mirabel et al. (1992). }
\enddata
\end{deluxetable}

\clearpage
\begin{deluxetable}{lcccc}
\tablecolumns{5}
\tablewidth{0pc}
\tablecaption{Possible Emission Lines from 1E1740.7-2942\tablenotemark{1}}
\tablehead{
Ion&\colhead{Rest-Frame}&\colhead{Measured}&\colhead{Blue-Shifted}&\colhead{Red-Shifted} \\
Name&$\lambda$ (\AA)\tablenotemark{2}&$\lambda$ (\AA)\tablenotemark{3}&$\lambda$ (\AA)\tablenotemark{3}&$\lambda$ (\AA)\tablenotemark{3}}
\startdata
Fe XXVI&1.78&1.78&1.70&1.96 \\
Ca XX&2.24&2.24&2.14&2.42 \\
 &2.54&2.54&2.44&2.68 \\
 &3.04&3.04&2.93&\nodata\tablenotemark{4} \\
Ca XIX&2.60&2.60&\nodata&\nodata \\
 &3.21&3.21&3.11&3.42 \\
S XVI&3.65&3.64&3.54&3.81 \\
 &4.73&4.71&4.61&4.90 \\
Ar XVII&3.95&3.95&3.86&4.14 \\
S XV&4.31&4.31&4.21&4.54 \\
 & 5.04&5.04&\nodata&\nodata \\
Si XVI&4.95&4.95&\nodata&\nodata \\
\tablenotetext{1}{All lines were detected only at a significance level of roughly $2 \sigma$. The equivalent width is typically $\lesssim$ 0.3\% of the wavelength of a line. }
\tablenotetext{2}{From Mewe et al. (1985).}
\tablenotetext{3}{The measurement accuracy is about 0.02 \AA, corresponding to the size of the MEG data bins.}
\tablenotetext{4}{The line seems to be blended with those of Ca XIX. }
\enddata
\end{deluxetable}

\clearpage
\begin{figure}
\psfig{figure=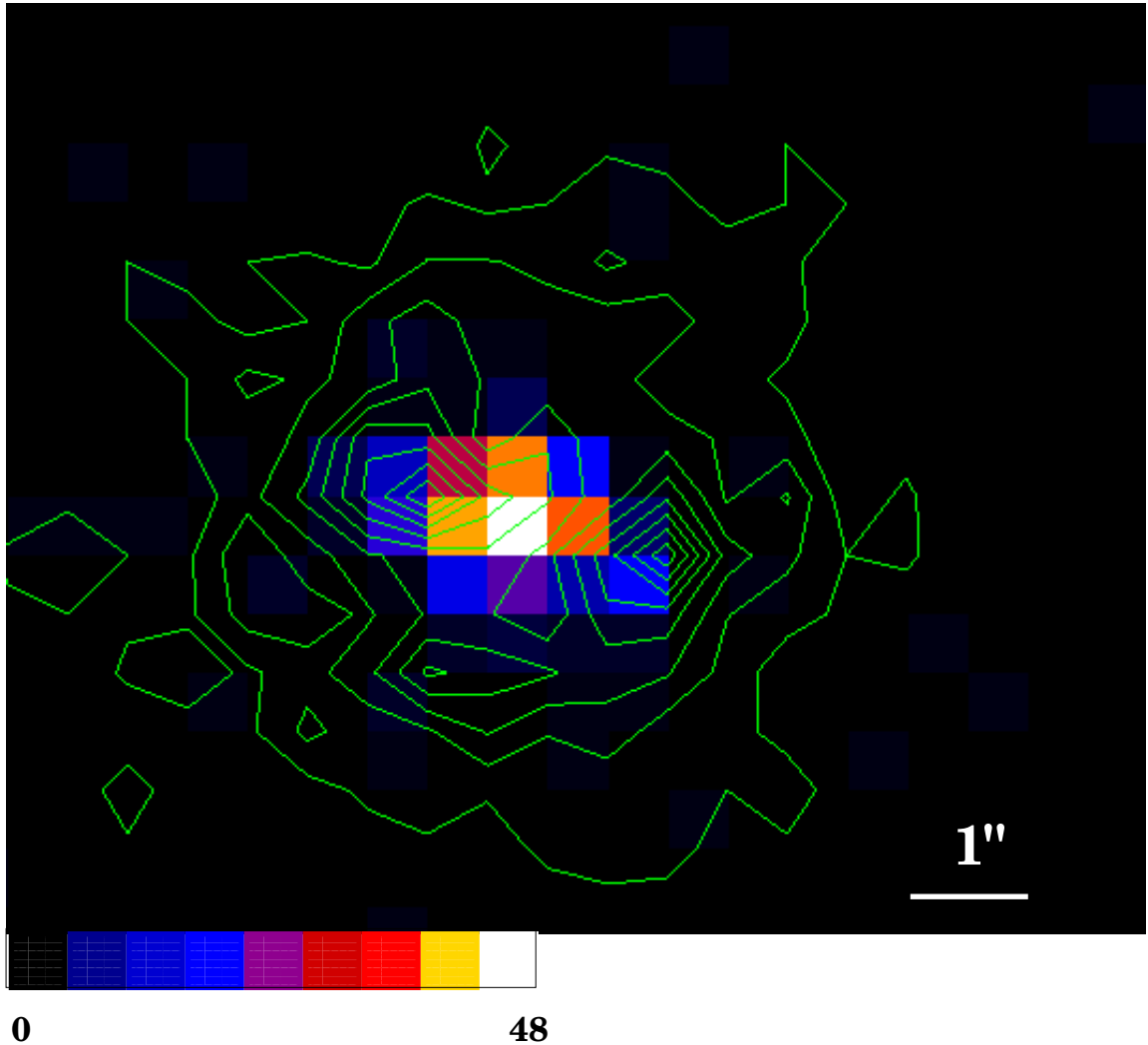,width=6in}
\caption{Unsmoothed X-ray image of 1E1740.7-2942. The image is made 
from the short-frame data (with pixel size 0.492\arcsec) and the 
overlaying contours are from the long-frame data. The image is color
coded linearly based on the observed X-ray intensity (in terms of
total counts). The color scheme is shown at the bottom. The contours 
are in the range of 5 to 43 counts, with a linear step size of 4.75 
counts. Note the ``lumpiness'' of the contours is an indication of 
significant pile-up in the long-frame data. }
\end{figure}

\begin{figure}
\psfig{figure=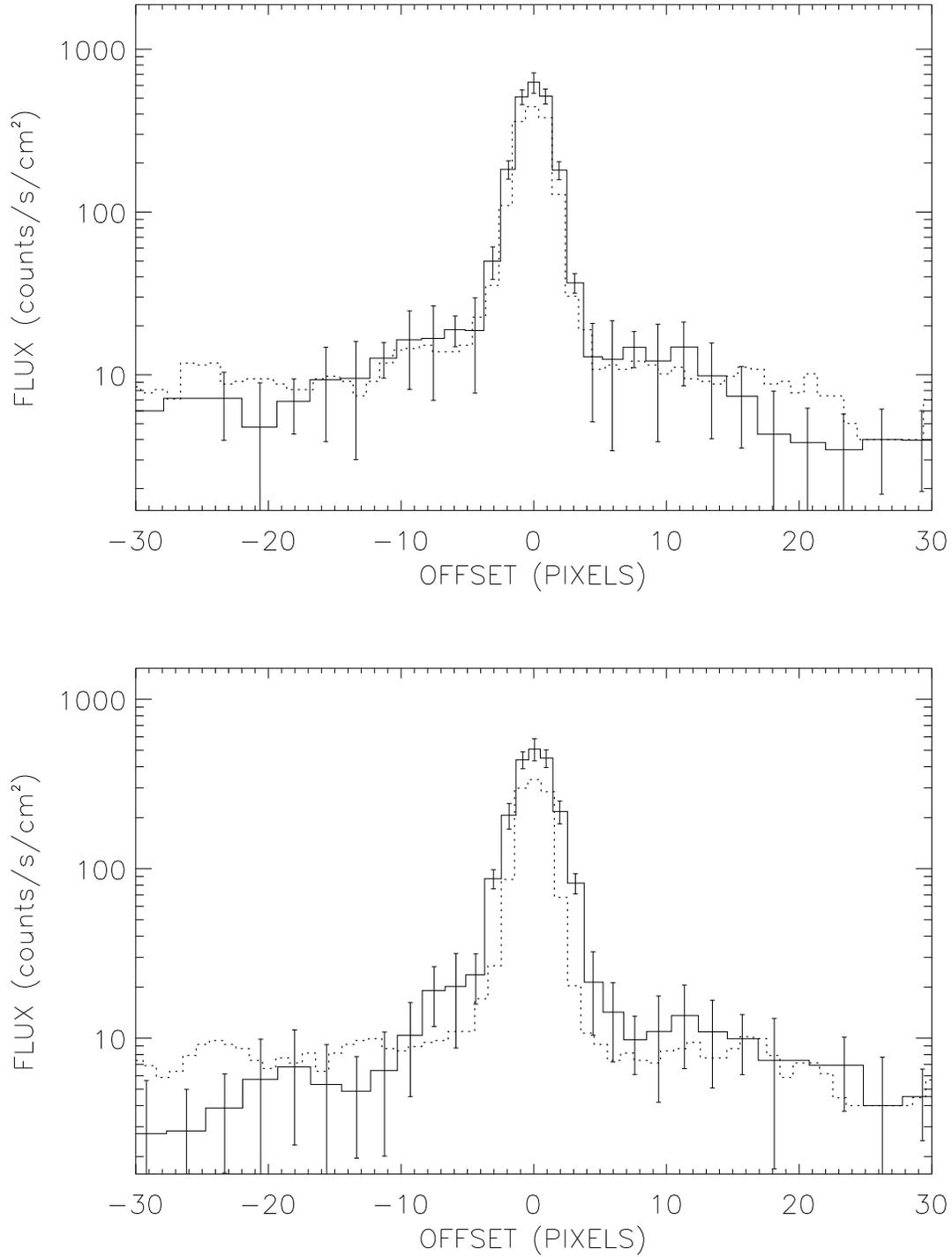,width=6in}
\caption{Linear profiles of the extended emission (in solid
histograms). The profiles were made by collapsing the observed 
counts along the direction of the elongation (bottom) and across it 
(top), respectively. The best-fit MARX simulated point-spread 
functions (PSFs) are also shown in dashed histograms. } 
\end{figure}

\begin{figure}
\psfig{figure=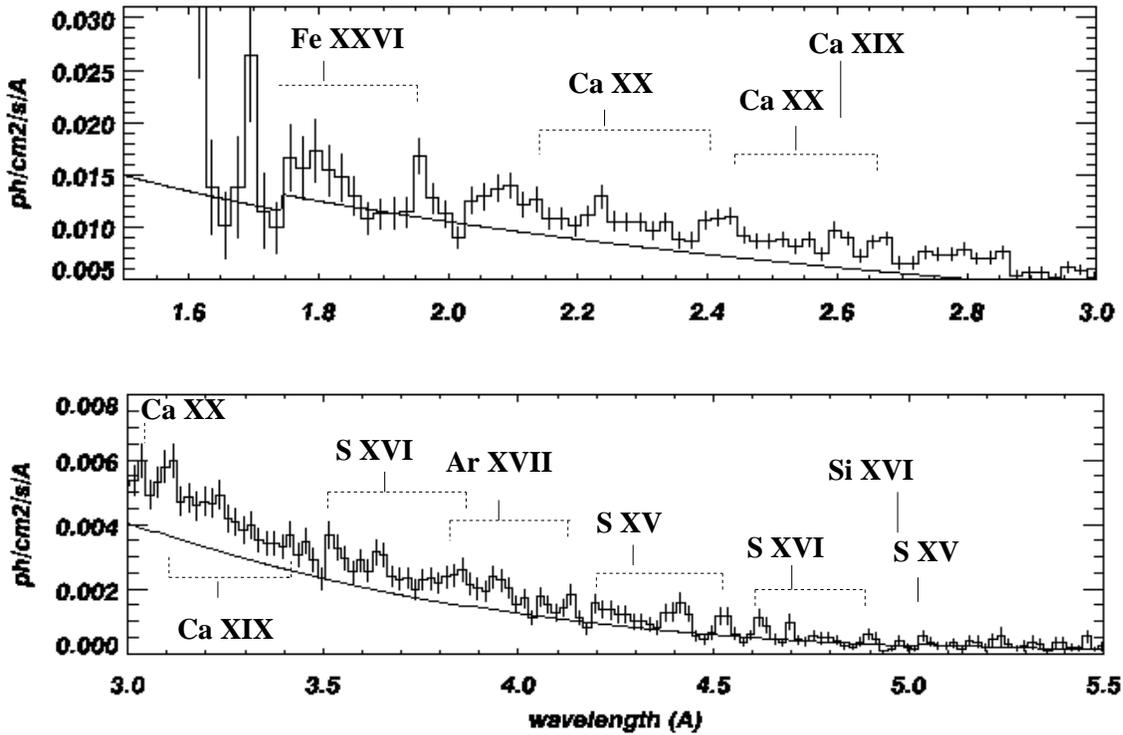,width=6in}
\caption{The first-order MEG spectrum of 1E1740.7-2942. The data is
binned to 0.02 \AA\ , which is roughly the MEG resolution. The
best-fit continuum model is also shown in solid line but is arbitrarily 
shifted down by 15\% to highlight possible emission lines. Each line 
candidate, as well as its proposed Doppler-shifted pairs, is marked 
with a tentitave identification (see text). }
\end{figure}

\begin{figure}
\psfig{figure=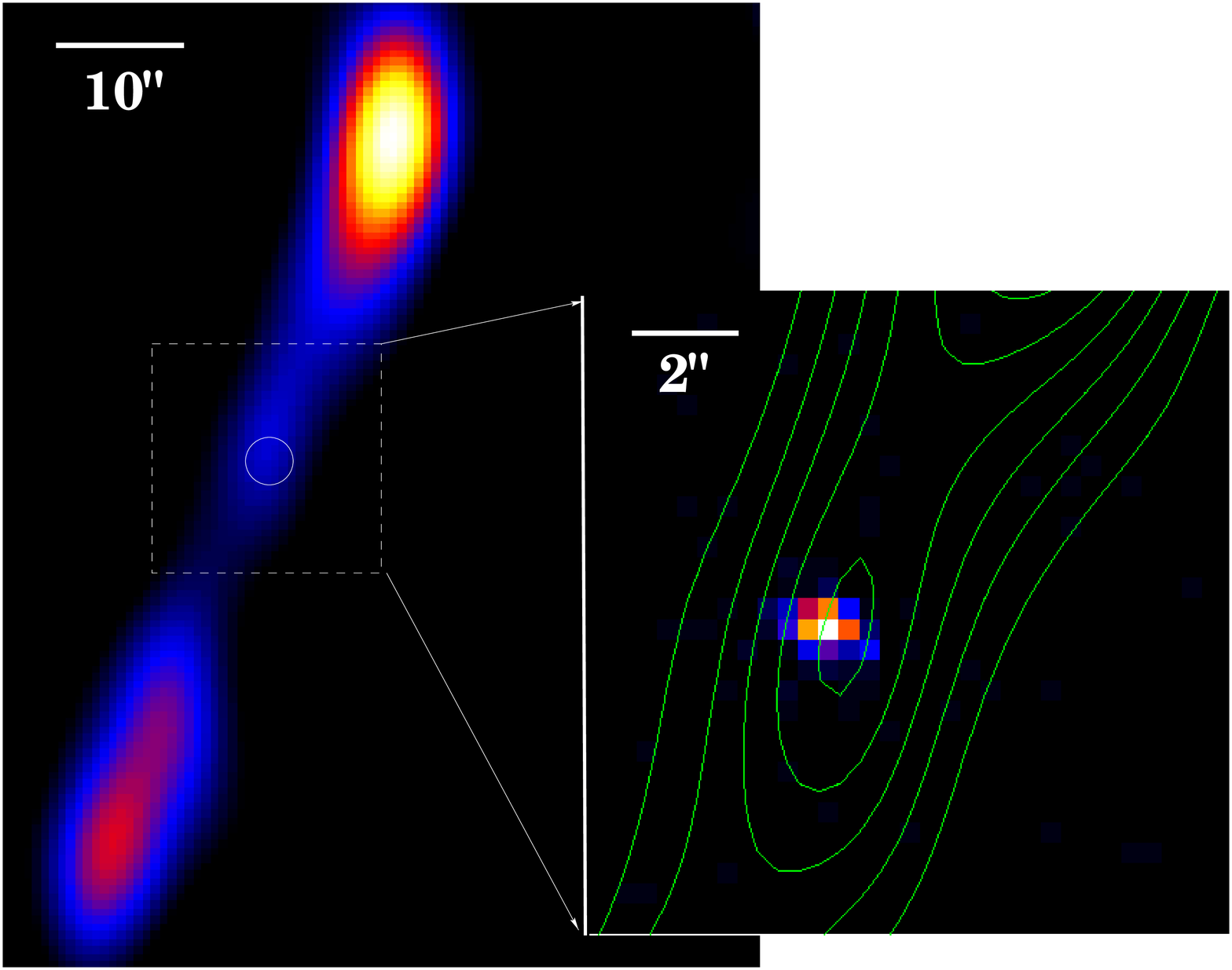,width=6in}
\caption{Comparison of X-ray and VLA radio (6-cm) images of
1E1740.7-2942. The
main panel shows the compact radio core and extended lobes (or jets)
and the inset shows the short-frame image overlaid by the radio
contours for a smaller central region (as indicated in the main
panel). The small circle in the main panel is indicative of the 
spatial extent of the X-ray emission observed. Note a small residual 
discrepancy between the position of the X-ray source and that of its
radio counterpart. }
\end{figure}


\clearpage
\begin{references}
\reference{} Allen,~C., et al. 1998, {\it SPIE}, 3444, 198
\reference{} Bally,~J., \& Leventhal,~M. 1991, Nature, 353, 234
\reference{} Bouchet,~L., et al. 1991, \apj, 383, L45
\reference{} Churazov,~E., Gilfanov,~M., \& Sunyaev,~R. 1996, \apjl,
464, L71
\reference{} Cook,~W.~R., Grunsfeld,~J.~M., Heindl,~W.~A.,
Palmer,~D.~M., Prince,~T.~A., Schindler,~S.~M., \& Stone,~E.~C., \apjl,
372, L75
\reference{} Harris,~M.~J. 1997, Proc. of the Fourth Compton
Symposium, Eds. C.~D.~Dermer, M.~S.~Strickman, \& J.~D.~Kurfess (New
York: AIP), p. 418
\reference{} Hertz,~P., \& Grindlay,~J.~E. 1984, \apj, 278, 137
\reference{} Kuznetsov,~S., et al. 1997, \mnras, 292, 651
\reference{} Leahy,~D.~A., Langill,~P., \& Kwok,~S. 1992, \aap, 259,
209 
\reference{} Lin,~D., Smith,~I.~A., Bottcher,~M. and
Liang,~E.~P. 2000, \apj, 531, 963
\reference{} Kotani,~T., Kawai,~N., Matsuoka,~M., \&
Brinkmann,~W. 1996, \pasj, 48, 619
\reference{} Marti,~J., et al. 2000, in preparation
\reference{} Mauche,~C.~W., \& Gorenstein,~P. 1986, \apj, 302, 371
\reference{} Mewe,~R., Gronenschild,~E.~H.~B.~M., \& van den Oord,~G.~H.~J. 
1985, \aaps, 62, 197
\reference{} Mirabel,~I.~F., Morris,~M., Wink,~J., Paul,~J., \&
Cordier,~B. 1991, \aap, 251, L43
\reference{} Mirabel,~I.~F., Rodr\'\i guez,~L.~F., Cordier,~B.,
Paul,~J., \& Lebrun,~F. 1992, Nature, 358, 215
\reference{} Mirabel,~I.~F., Rodr\'\i guez,~L.~F., Cordier,~B.,
Paul,~J., \& Lebrun,~F. 1993, \aaps, 97, 193
\reference{} Osterbrock,~D.~E. 1989, Astrophysics of Gaseous Nebulae
and Active Galactic Nuclei, Mill Valley: University Science Books
\reference{} Paragi,~Z., Vermeulen,~R.~C., Fejes,~I.,
Schilizzi,~R.~T., Spencer,~R.~E., \& Stirling,~A.~M. 2000, \aap, in
press. 
\reference{} Predehl,~P., \& Schmitt,~J.~H.~M.~M. 1995, \aap, 293, 889
\reference{} Phillips,~J.~A., Joseph,~T., \& Lazio,~W. 1995, \apj,
442, L37
\reference{} Sakano,~M., Imanishi,~K., Tsujimoto,~M, Koyama,~K., \&
Maeda,~Y. 1999, \apj, 520, 316
\reference{} Sheth,~S., Liang,~E., Luo,~C., \& Murakami,~T. 1996,
\apj, 468, 755
\reference{} Skinner,~G.~K., et al. 1987, Nature, 330, 544
\reference{} Skinner,~G.~K., et al. 1991, \aap, 252, 172
\reference{} Smith,~D.~M., Heindl,~W.~A., Swank,~J., Leventhal,~M., 
Mirabel,~I.~F., \& Rodr\'\i guez,~L.~F. 1997, \apj, 489, L51
\reference{} Sunyaev,~R., et al. 1991, \apj, 383, L49
\reference{} van Speybroeck,~L., Jerius,~D., Edgar,~R.~J.,
Gaetz,~T.~J., Zhao,~P., \& Reid,~P.~B. 1997, Proc. SPIE, 3113, 98
\reference{} Weisskopf,~M.~C., \& O'Dell,~S.~L. 1997, Proc. SPIE,
3113, 2
\reference{} Wise,~M., et al. 1999, MARX 2.0 User's Guide
\reference{} Yan,~M., \& Dalgarno,~A. 1997, \apj, 481, 296
\end{references}
\end{document}